\documentclass[superscriptaddress,aps,physrev,preprint,groupedaddress]{revtex4-2}
\usepackage{graphicx}
\usepackage{dcolumn}
\usepackage{bm}
\usepackage{physics}
\usepackage{amsmath,multirow}
\usepackage[colorlinks=false,citecolor=blue]{hyperref}
\graphicspath{Figures/}
\usepackage{booktabs}

\let\oldCite\cite
\renewcommand*{\cite}{~\oldCite}

\begin{document}

\title{Theoretical characterization of NV-like defects in 4H-SiC using ADAQ with the SCAN and r$^2$SCAN meta-GGA functionals}

\author{Ghulam Abbas}
 \email{Corresponding authors: ghulam.abbas@liu.se}
\affiliation{Department of Physics, Chemistry and Biology, Link\"oping
  University, SE-581 83 Link\"oping, Sweden}

\author{Oscar Bulancea-Lindvall} 
\affiliation{Department of Physics, Chemistry and Biology, Link\"oping
  University, SE-581 83 Link\"oping, Sweden}

\author{Joel Davidsson} 
\affiliation{Department of Physics, Chemistry and Biology, Link\"oping
  University, SE-581 83 Link\"oping, Sweden}

\author{ Rickard Armiento}
 \email{rickard.armiento@liu.se}   
\affiliation{Department of Physics, Chemistry and Biology, Link\"oping
  University, SE-581 83 Link\"oping, Sweden}

\author{Igor A. Abrikosov}
 \email{igor.abrikosov@liu.se}
\affiliation{Department of Physics, Chemistry and Biology, Link\"oping
  University, SE-581 83 Link\"oping, Sweden}
\date{\today}

\begin{abstract}
    Kohn-Sham density functional theory (DFT) is widely used for screening color centers in semiconductors. While the Perdew-Burke-Ernzerhof (PBE) functional is efficient, it often lacks precision in describing defects. The Heyd-Scuseria-Ernzerhof (HSE) functional is more accurate but computationally expensive, making it impractical for large-scale screening. However, third-rung functionals of ``Jacob's ladder'' remain largely underexplored in this context. This study evaluates the Strongly Constrained and Appropriately Normed (SCAN) family of meta-GGA functionals as potential alternatives to PBE for characterizing NV-like color centers in 4H-SiC using the Automatic Defect Analysis and Qualification (ADAQ) framework. We examine nitrogen, oxygen, fluorine, sulfur, and chlorine vacancies in 4H-SiC, focusing on applications in quantum technology. Our results show that SCAN and r²SCAN achieve greater accuracy than PBE, approaching HSE’s precision at a lower computational cost. This suggests that the SCAN family offers a practical improvement for screening new color centers, with computational demands similar to PBE.
\end{abstract}

\maketitle

\newpage

Point defects in wide-band-gap semiconductors, such as nitrogen-vacancy (NV) centers in diamond, exhibit optically addressable localized electronic states that enable transformative applications in quantum computing\cite{1}, sensing\cite{2,3}, and nanothermometry\cite{4}. The remarkable success of NV centers has inspired a growing search for novel defects in alternative host materials\cite{5,6,7,8,9,10,11,12}, with Kohn-Sham density functional theory (DFT) playing a pivotal role in their theoretical characterization\cite{13,13_2} and high-throughput screening\cite{5,6,9,10,11,12,14,15,16}. This effort is further advanced by frameworks like the Automatic Defect Analysis and Qualification (ADAQ)\cite{14}, which automate magneto-optical property simulations to expedite the discovery of promising quantum defects\cite{9,10,11,12}.

The accuracy of KS-DFT largely depends on the exchange-correlation energy functional $E_\text{xc}[\rho]$, for which various approximations have been developed since the 1960s\cite{13,19}. These approximations differ in computational efficiency and the complexity of their mathematical formulations\cite{17,18,22}. In Perdew and Schmidt's “Jacob’s ladder” classification\cite{21}, semi-local functionals are arranged in increasing levels of sophistication and accuracy along the first three rungs. Such functionals are conveniently expressed as a single integral:
\begin{equation} E_\text{xc} = \int \epsilon_\text{xc}(\vec{r})d^3\vec{r}, \end{equation}
where $\epsilon_\text{xc}(\vec{r})$ is the exchange–correlation energy density per unit volume. On the first rung, local density approximations (LDA)\cite{13,17} define $\epsilon_\text{xc}(\vec{r})$ solely as a function of the electron density $\rho(\vec{r})$. On the second rung, generalized gradient approximations (GGA)\cite{18} incorporate both $\rho(\vec{r})$ and its gradient $\nabla\rho(\vec{r})$. On the third rung, meta-GGA functionals\cite{22} extend $\epsilon_\text{xc}(\vec{r})$ to include $\rho(\vec{r})$, $\nabla\rho(\vec{r})$, the kinetic-energy density $t = \frac{1}{2}\sum_i \nabla\phi^* \nabla\phi$, and/or the Laplacian $\nabla^2\rho$.

The GGAs, like the Perdew–Burke–Ernzerhof (PBE) functional\cite{18,23}, are widely used for their balance of accuracy and computational efficiency, particularly in large-cell systems, high-throughput screenings, and ab initio molecular dynamics. However, PBE suffers from self-interaction errors\cite{24}, fails to capture medium- and long-range dispersion\cite{25}, and systematically underbinds\cite{26}, leading to overestimated lattice parameters\cite{27}. Additionally, PBE significantly underestimates KS bandgaps compared to experiments and higher-order methods\cite{54}. To improve KS-based DFT beyond GGAs, exact-exchange potentials from Hartree-Fock (HF) theory, expressed as a double integral over the occupied orbitals, can be used. These potentials involve either a full Coulomb potential, as in Becke 3-parameter Lee–Yang–Parr fucntional (B3LYP)\cite{55} and PBE0\cite{61}, or a screened potential, as in Heyd–Scuseria–Ernzerhof (HSE)\cite{29,30} and Yukawa-screened PBE0. While hybrid DFT functionals (fourth rung of Jacob’s Ladder) offer greater accuracy, their higher computational cost renders them impractical for high-throughput screening.

To overcome the limitations of LDA and GGA, meta-GGAs have gained popularity for computing material properties. Among these, the Strongly Constrained and Appropriately Normed (SCAN)\cite{31} family of meta-GGAs has emerged as a reliable choice in DFT calculations, offering improved accuracy in lattice parameters and solid-state energetics\cite{58,22,32,53,33,34}. The SCAN adheres to all known semi-local functional constraints and approximates chemical bonding strengths comparable or even better than hybrid DFT. However, numerical instabilities in the original SCAN for certain systems led to regularization efforts\cite{26,35,36}, which relaxed some constraints. However, the latest variant, r$^2$SCAN, has been designed to achieve numerical
efficiency along with transferable accuracy of SCAN while satisfying all constraints\cite{37}.

Due to these developments, it timely to reassess the potential of meta-GGAs, particularly SCAN, for modeling point defects in wide-band-gap semiconductors and evaluating their suitability for high-throughput calculations. To establish SCAN's advantages over PBE, it is essential to examine its accuracy in predicting defect formation energies, charge transition levels (CTLs), and atomic relaxations compared to experimental data and high-level benchmarks. While SCAN has shown improved potential energy surfaces for diamond color centers\cite{38}, its application to defect excited states using $\Delta$SCF\cite{42,43} in high-throughput setups remains unexplored.

This study evaluates the SCAN and r$^2$SCAN meta-GGA functionals for defect formation energies and optical transition levels in 4H-SiC, a wide-band-gap semiconductor. Using ADAQ workflows\cite{14} with VASP\cite{39}, we benchmark SCAN functionals for ab initio screening of NV-like color centers, including nitrogen\cite{6}, oxygen\cite{8}, fluorine, sulfur, and chlorine vacancy clusters\cite{12}, which are promising for quantum technologies. SCAN results are compared to HSE06 hybrid functional and experimental data when available. SCAN achieves optical parameters closer to HSE at a fraction of the computational cost, making it a cost-effective alternative for high-throughput screening of color centers in solids.

\begin{figure}
    \centering
     \includegraphics[width=1\linewidth]{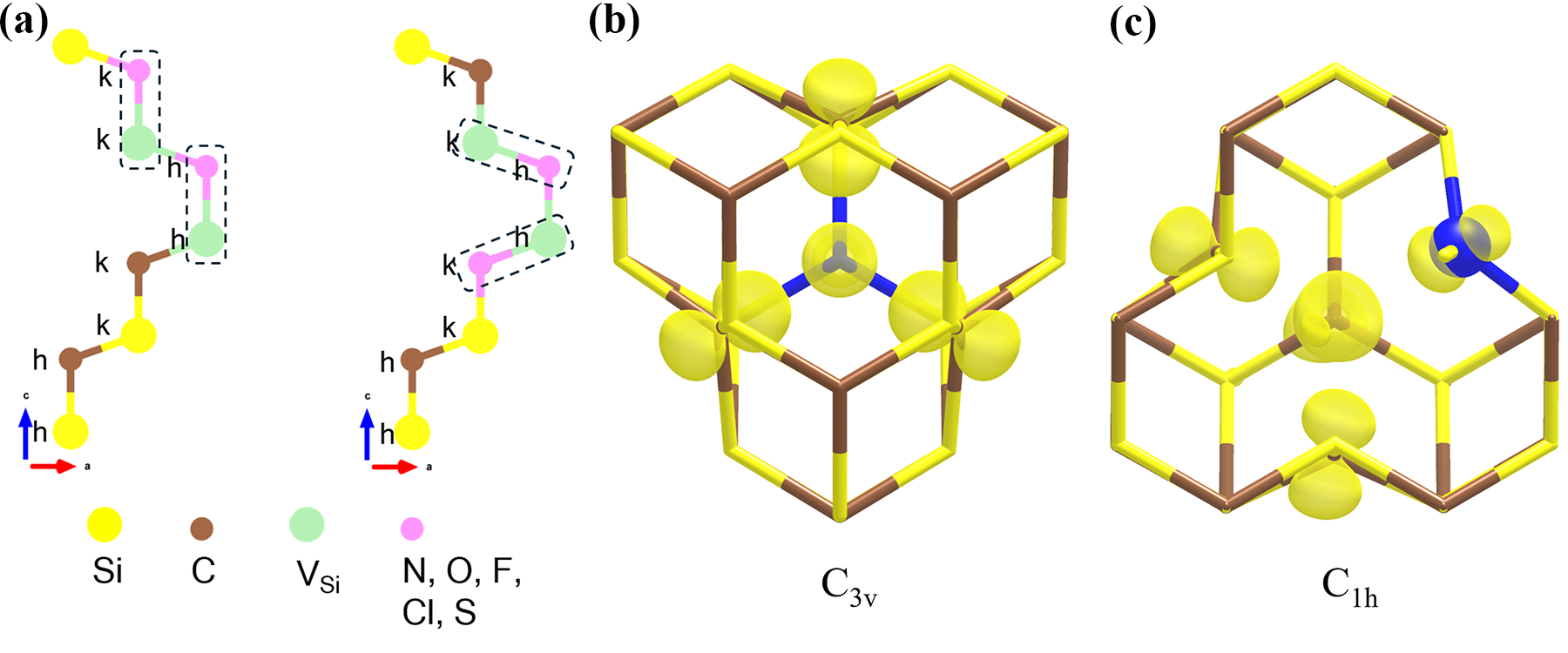}
\caption{(a) Four NV-like defect configurations in 4H-SiC, where a heteroatom substitutes a C and a Si vacancy is present. Axial defects are labeled \textit{hh} and \textit{kk}, and basal defects are \textit{hk} and \textit{kh}. The “\textit{h}” (quasihexagonal) and “\textit{k}” (quasicubic) refer to nonequivalent lattice sites. Defect-local structures of NV$^{-1}$ centers in (a) \textit{hh} and (b) \textit{kh} configurations, showing the occupied a$_1$ and a$^\prime$ states with $0.005$~bohr$^{-3}$ spin density isosurfaces.}
    \label{fig:1}
\end{figure}

Point-defect characterization calculations are systematically automated using workflows in ADAQ\cite{14} to assess the SCAN and r$^2$SCAN functionals. These workflows, implemented with the high-throughput toolkit (\emph{httk})\cite{40,60}, handle tasks from defect generation to magneto-optical property characterization. Calculations employ spin-polarized DFT simulations with the PAW method in the Vienna Ab-initio Simulation Package (VASP)\cite{39,41}, using SCAN\cite{31} and its regularized form r$^2$SCAN\cite{37} as exchange-correlation functionals. The defects for $p$-elements (nitrogen, oxygen, fluorine, sulfur, chlorine) were generated in a double-defect configuration, similar to the NV center(Figure \ref{fig:1}a), using a 576-atom (6×6×2) supercell. Th SCAN and r$^2$SCAN calculations were performed at the $\Gamma$-point, ensuring energy convergence below $10^{-4}$eV and interatomic forces under $0.005$eV\AA$^{-1}$, in line with high-throughput screening workflows. For HSE06 functional calculations a slightly lower force threshold of 0.01~eV\AA$^{-1}$ is used.

To begin, we calculated and compared the lattice parameters, bulk modulus, and electronic band gaps of 4H-SiC using PBE, SCAN, r²SCAN, and HSE functionals, with results shown in Table~\ref{tab:1}. To align with theoretical values, a correction of 0.02$\%$ was applied to the experimental lattice constants to compensate for anharmonicity of zero-point vibrations, yielding $a = 3.071$~\AA\cite{46}. The PBE overestimates lattice constants and underestimates the bulk modulus by $7$~GPa. SCAN and r²SCAN closely match the corrected lattice constants, with slight bulk modulus overestimation of $6–7$~GPa. HSE underestimates lattice constants and overestimates the bulk modulus by $13$~GPa. Thus, SCAN shows the best agreement with experimental values. Additional comparisons of elastic, Young’s, and shear moduli are provided in Table S1. Overall, meta-GGAs offer a nearly ideal description of 4H-SiC properties, while PBE predicts a softer structure and HSE a harder one.

\begin{table}
\centering
\caption{Calculated lattice constants (Å), bulk moduli $B$ (GPa), band gaps $E_g$ (eV), and band edge shifts $\Delta E_{\text{VBM}}$, $\Delta E_{\text{CBM}}$ (eV) with respect to the HSE band edges of 4H-SiC. A comparison of the relative computational time (RCT) percentage of semi-local approximations relative to HSE bandgap calculations is also provided.}.

\begin{ruledtabular}
\begin{tabular}{>{\hspace{0pt}}m{0.112\linewidth}>{\hspace{0pt}}m{0.290\linewidth}>{\hspace{0pt}}m{0.106\linewidth}>{\hspace{0pt}}m{0.16\linewidth}>{\hspace{0pt}}m{0.2\linewidth}>{\hspace{0pt}}m{0.07\linewidth}} 
\textbf{Method}          & \textbf{Lattice Constants} \par{}\textbf{(\AA)}                                          & \textbf{\( B \)}\par{}\textbf{(GPa)} & \textbf{$E_g$}\par{}\textbf{(eV)} & \textbf{\textbf{$\Delta E_{\text{VBM}}$, $\Delta E_{\text{CBM}}$} (eV)} & \textbf{RCT} \par{}\textbf{(\% )}  \\ 
\hline
\multirow{2}{*}{Exptl.} & \textit{a} = 3.078\cite{47} (3.071\footnote{\mbox{Zero Point Correction.}}) & \multirow{2}{*}{224\cite{48}}                        & \multirow{2}{*}{3.23}                           & \multirow{2}{*}{\textemdash}                                & \multirow{2}{*}{\textemdash}\\
 & \par{}\textit{c} = 10.085 &    &       &                  &   \\
PBE                      & \textit{a} = 3.090, \textit{c} = 10.117                                      & 217                            & 2.22 (3.16\footnote{\mbox{With HSE bandgap calculated at theoratical lattice parameters for the corresponding functional.}}) & 0.91, -0.05                      & 2.4                               \\
SCAN                     & \textit{a} = 3.071, c = 10.057                                               & 230                            & 2.60 (3.17\textsuperscript{b}) & 0.51, -0.07                      & 2.8                               \\
r\textsuperscript{2}SCAN & \textit{a} = 3.074, \textit{c} = 10.066                                      & 231                            & 2.61                           & 0.50, -0.07                       & 2.6                               \\ 
HSE06                    & \textit{a} = 3.069, \textit{c} = 10.042                                      & 237                            & 3.18                           & \textemdash                                & 100                               \\

\end{tabular}
\end{ruledtabular}
 \label{tab:1}
\end{table}

The most notable difference among the functionals lies in the electronic band gap (Table~\ref{tab:1}). The PBE functional underestimates the Kohn-Sham bandgap of 4H-SiC by $1.1$~eV compared to experimental values. SCAN and r${^2}$SCAN functionals offer modest improvements, reducing the deviation to $0.7$~eV. The HSE fucntional provides the closest match, with a difference of only $0.05$~eV. Additionally, using HSE functional to calculate band gaps for geometries relaxed with the PBE or SCAN's yields values close to experimental results (Table\ref{tab:1}). The final column of Table\ref{tab:1} compares computational costs for a single static self-consistent calculation: SCAN and r$^2$SCAN functionals require only 2.8$\%$ and 2.6$\%$, respectively, of the computational time for HSE, slightly more than PBE’s 2.4$\%$. All calculations were performed with identical parallelization settings, demonstrating the efficiency of SCAN and r$^2$SCAN relative to HSE. We also compare band gap edges, i.e., conductance band minima (CBM) and valence band maxima (VBM), aligned on an absolute energy scale relative to the vacuum level [52, 53]. This alignment, crucial for comparing charge transition levels (CTLs), shows band edge shifts relative to HSE in Table I. The band gap does not close symmetrically: VBM shifts by $0.51$~eV (SCAN) and $0.91$~eV (PBE), while CBM shifts slightly by $-0.07$~eV (SCAN) and $-0.05$~eV (PBE). While PBE functional is computationally efficient, the SCAN family functionals provides crystal structure parameters closer to experiments with moderately improved bandgap descriptions.

\begin{figure}
    \centering
     \includegraphics[width=1\linewidth]{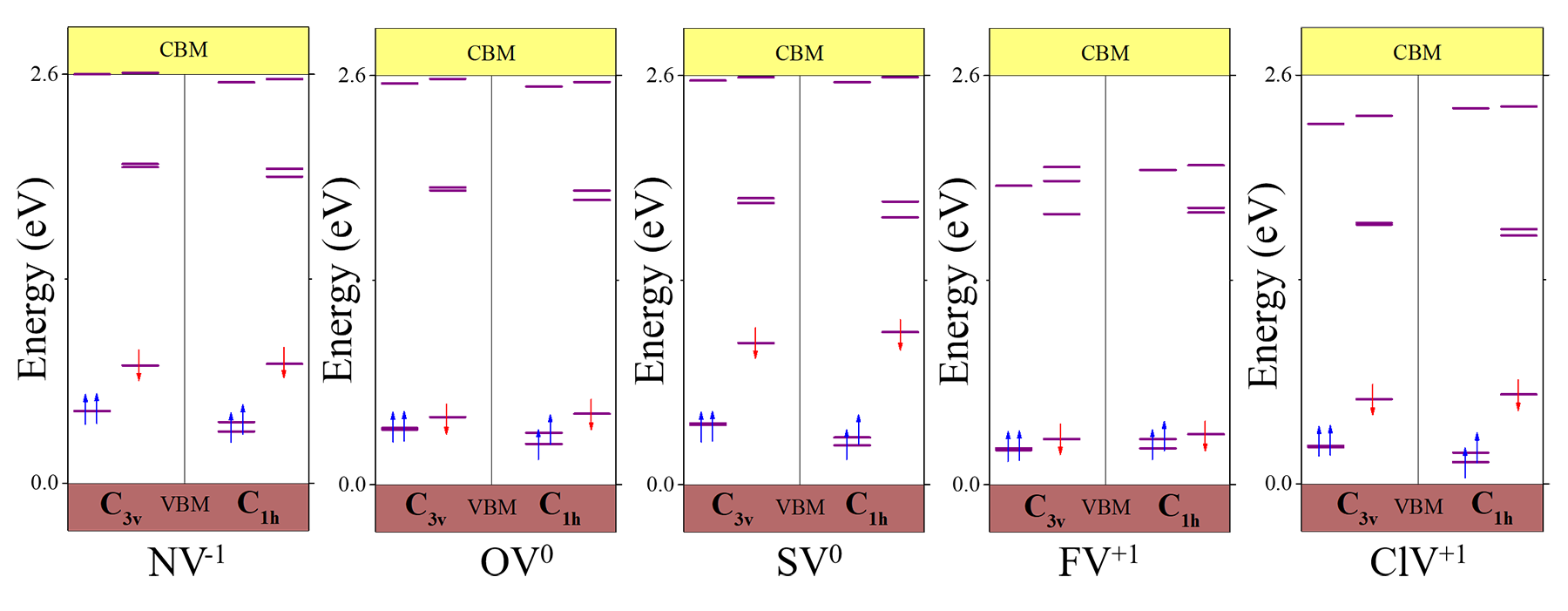}
    \caption{Kohn-Sham eigenstates of NV-like defects in 4H-SiC calculated with the SCAN functional for axial configurations (degenerate defect levels due to C\textsubscript{3v} symmetry) and basal configurations (lifting of degeneracy due to C\textsubscript{1h} symmetry). Blue and red arrows indicate electron spins in the ground states.}
    \label{fig:2}
\end{figure}

The studied NV-like defects (NV$^{-1}$, OV$^{0}$, SV$^{0}$, FV$^{+1}$, and ClV$^{+1}$) consist of a silicon vacancy paired with a substitutional heteroatom ion on a neighboring carbon site, forming double defect clusters in 4H-SiC. These clusters can exist in four nonequivalent lattice configurations—\textit{kk}, \textit{hk}, \textit{kh}, and \textit{hh}—with \textit{kk} and \textit{hh} along the \textit{c}-axis (C$_{\text{3v}}$ symmetry), and \textit{kh} and \textit{hk} off-axis (C$_{\text{1h}}$ symmetry) as shown in Figure~\ref{fig:1}a. The spin densities for the highest occupied a$_{1}$(a$^\prime$) states of NV$^{-1}$ in \textit{hh} and \textit{kh} configurations are illustrated in Figure~\ref{fig:1}b, c. From group theory analysis, the double defect clusters create multiple localized states within the bandgap. The SCAN-calculated quasi single-particle diagrams (Figure~\ref{fig:2}) show defect states within the bandgap, including lower $e$ states and higher a$_{1}$ states in the spin-up channel, with an additional a$_{1}$ (a$^\prime$) state in the spin-down channel. The electronic configurations of stable high-spin states are shown for C$_\text{3v}$ and C$_\text{1h}$ symmetries, where C$_\text{3v}$ configurations feature doubly degenerate $e$ states, while this degeneracy is lifted to produce a$^\prime$ and a$^{\prime\prime}$ states in C$_{\text{1h}}$. To mimic the electronic occupancy of NV$^{-1}$, OV and SV defects remain neutral, while FV and ClV adopt single positive charge, stable  triplet states, consistent with their positions in the periodic table. Despite SCAN functional bandgap underestimation, it produces localized states within the bandgap in close resemblance to HSE results\cite{6,8,12}, as detailed in Figure~\ref{fig:2}.

Three different charge states with different spin configurations were compared through examination of formation energy calculations by equation\cite{59}:
\begin{equation}
    E^{q,S}_\text{form}(E_F) = E^{q,S}_\text{defect} - E_\text{bulk} - \sum_i n_i\mu_i + q(\varepsilon_\text{F}+E_\text{V}) + E_\text{corr}^q\text{,}
\end{equation}
where $E^{q,S}_\text{defect}$ denotes the total energy of the defect state of with charge $q$ and spin state $S$, while $E_\text{bulk}$ is the energy of the defect-free host system, and $n_i$ and $\mu_i$ denote the changes in the chemical components to form the defect and respective chemical potential. The chemical potentials are here taken in the rich abundance limit, assuming the 0 K crystalline structures of each element, and $\varepsilon_\text{F}$ is Fermi energy (measured from top of the theoretical valance band maxima, $E_\text{V}$). Finally a charge correction term $E_\text{corr}^q$ to account for finite-size coulombic effects. In ADAQ, the Lany-Zunger (LZ)\cite{28} charge correction scheme is implemented, that gives the same correction for all defects. 

The point of $\varepsilon_\text{F}$ at which states \textit{$q$} and \textit{$q^\prime$} can exist with the same $E_\text{form}$ is the charge-state transition level (CTL), $\varepsilon(q/q^\prime)$, given by\cite{13_2}:
\begin{equation}
    \varepsilon(q/q^\prime)=\frac{\Delta{E_\text{form}}(q:\varepsilon_\text{F}=0)-\Delta{E_\text{form}}(q':\varepsilon_\text{F}=0)}{q^\prime-q},
\end{equation}
The CTL can be compared for different exchange-correlation functionals if their levels are aligned against a common reference, such as the average electrostatic potential or vacuum level. Similar to $\varepsilon_\text{F}$, $\varepsilon(q/q^\prime)$ is referenced with respect to the $E_\text{V}$. To reference the $E_\text{V}$ position relative to the vacuum level, we construct a 4H-SiC slab (001) model of 12 layers with hydrogen termination (for more details see Figure S1 in supplementary material). Initially, the ionic relaxation was carried out at the PBE level, and later, for the SCAN, r$^2$SCAN and HSE functionals, we scaled the lattice parameters to equilibrium constants for each level functional.

\begin{figure}[htbp]
    \centering
     \includegraphics[width=0.75\linewidth]{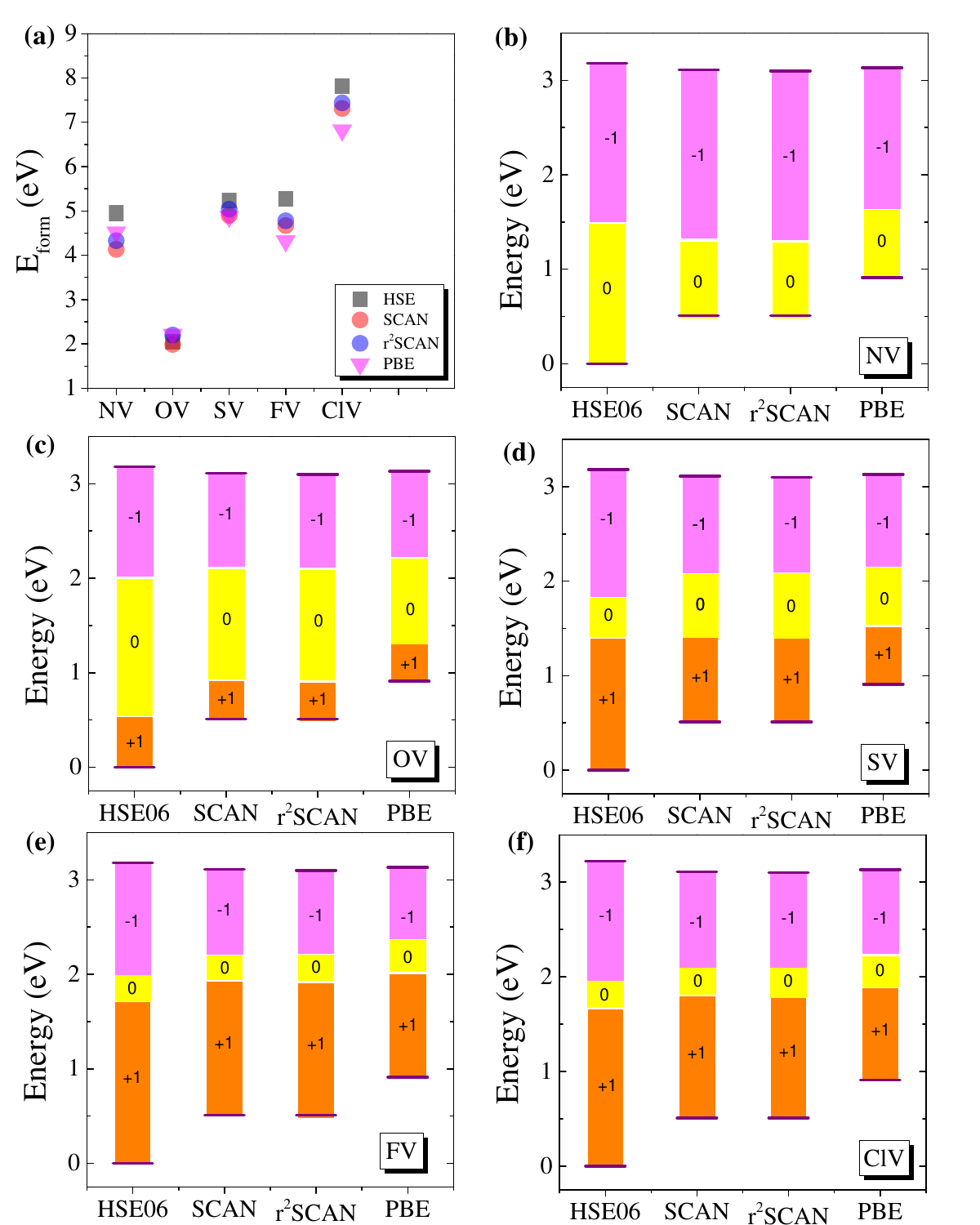}
     \caption{(a) Formation energy of the neutral charge lowest energy spin states. The HSE results for oxygen and chlorine are taken from the results already published elsewhere\cite{8,12}. The (b$-$f) charge transition level alignment (CTL) diagrams for NV-like defects in 4H-SiC. Horizontal bars correspond to the position of VBM and CBM for a given functionals.}
    \label{fig:3}
\end{figure}

We first compare $E^{q,S}_\text{form}$ of neutral defects ($q=0$), independent of $\varepsilon_\text{F}$. These formation energies, shown in Figure~\ref{fig:3}a, enable direct comparison of defect stability across functionals. In general, SCAN and r${^2}$SCAN yield formation energies between PBE and HSE results, except for NV. For NV in 4H-SiC, HSE gives the highest energy ($4.97$~eV) compared to PBE ($4.52$~eV), SCAN ($4.13$~eV), and r${^2}$SCAN ($4.33$~eV). In contrast, OV shows smaller differences, with SCAN and r${^2}$SCAN close to HSE ($1.98$~eV) and PBE ($2.21$~eV). The SV displays higher values: $4.87$~eV (PBE), $4.90$~eV (SCAN), $5.04$~eV (r${^2}$SCAN) and $5.23$~eV (HSE). For FV, formation energies vary more widely, with SCAN ($4.67$~eV) and r${^2}$SCAN ($4.78$~eV) between PBE ($4.32$~eV) and HSE ($5.23$~eV). Similarly, ClV exhibits a wider range: $6.82$~eV (PBE), $7.13$~eV (SCAN), $7.44$~eV (r${^2}$SCAN), and $7.82$~eV (HSE), with a notable ~1 eV variability between different functionals. These values, tied to the chemical environment, reflect the bond formation/disruption conditions in wide-bandgap semiconductors, where higher energies imply more extreme conditions. Formation energy trends across elements (e.g., O to S, F to Cl) reveal stepwise increases, offering further insight into defect comparisons.

This trend becomes more intriguing when examining CTL shifts for different defects. Figure~\ref{fig:3}b-f shows calculated CTLs for five defects using semi-local and hybrid functionals, referenced to vacuum as described. For comparison, only ($+1/0$) and ($0/{-1}$) CTLs for the ground-state configuration (lowest on the defect hull\cite{9}) are shown. Formation energy versus Fermi energy is provided in Figures S3 and S4 (supplementary material). Defect CTLs fall within 4H-SiC bandgaps, showing consistency across functionals despite varying bandgap estimates. The NV defect in 4H-SiC exhibits a CTL of ($0/{-1}$) only within the HSE-calculated bandgap, while semi-local approximations capture CTLs within reduced bandgaps. SCAN ($-0.19$~eV) and r$^2$SCAN ($0.20$~eV) functionals deviate more from HSE than PBE ($0.14$~eV). For four NV-like defects, we take HSE as a benchmark for accurately reproducing experimental bandgaps. The SCAN and r${^2}$SCAN align better with HSE (Mean absolute error (MAE): $0.19$~eV, $0.18$~eV) than PBE (MAE: $0.28$~eV), offering more accurate CTL predictions when using vacuum as common reference.

\begin{figure}
    \centering
     \includegraphics[width=1\linewidth]{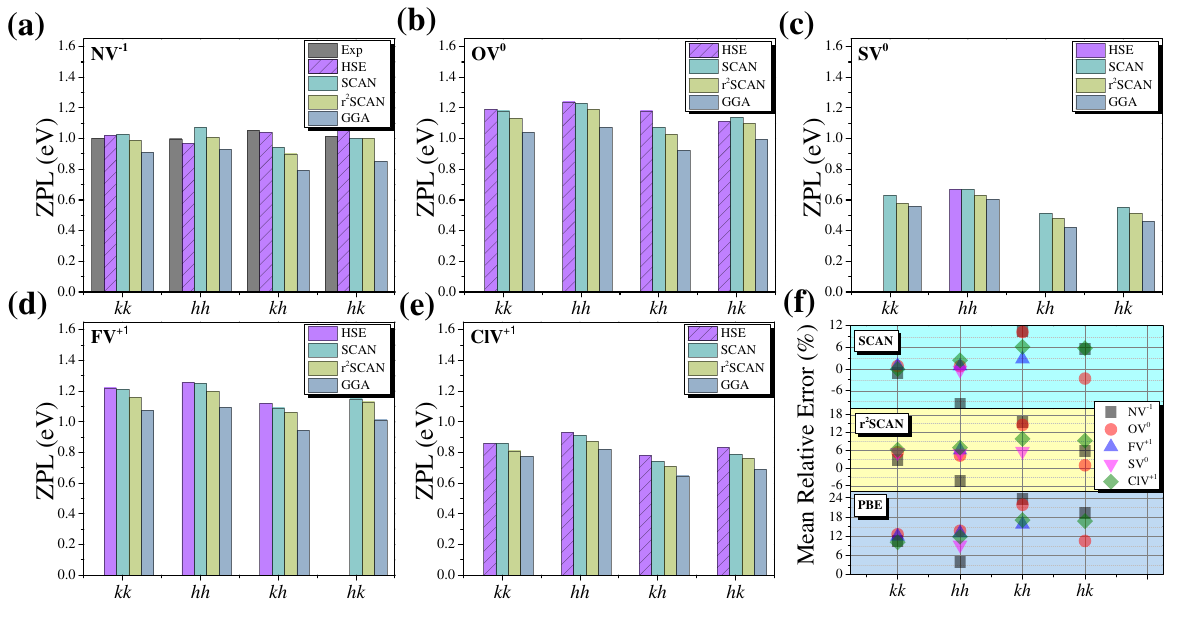}
   \caption{(a-e) Comparison of theoretical and experimental ZPL energies for NV-like defects in 4H-SiC. (f) Mean relative error percentages of SCAN, r$^2$SCAN, and PBE ZPL values compared to HSE. The HSE ZPL values for NV$^{-1}$\cite{6}, OV$^{0}$\cite{8}, and ClV$^{+1}$\cite{12} are taken from references.}
    \label{fig:4}
\end{figure}

To simulate optically addressable, excited triplet states in NV$^{-1}$, OV$^{0}$, SV$^{0}$, FV$^{+1}$, and ClV$^{+1}$ defects, the $\Delta$SCF\cite{42,43} approximation was applied using the automated screening workflow of ADAQ\cite{14}. Specifically, for on- (off-) axis defect configurations, an electron from the spin minority channel a$_{1}$(a$^\prime$) state is promoted to the e(a$^\prime$) level. Figure~\ref{fig:4} compares experimental and calculated ZPLs using various functionals. The HSE functional consistently yields the highest ZPL values, followed by SCAN, r$^2$SCAN, and PBE. However, NV$^{-1}$ shows exceptions: for \textit{kk}-NV$^{-1}$, r$^2$SCAN provides ZPLs closest to experiment\cite{6}, whereas HSE and SCAN overestimate, and PBE significantly underestimates. For \textit{hh}-NV$^{-1}$, SCAN and r$^2$SCAN overestimate, while HSE and PBE underestimate. In \textit{kh}-NV$^{-1}$, all functionals underestimate, whereas for \textit{hk}-NV$^{-1}$, only HSE overestimates. Despite these variations, HSE is widely recognized for accurate ZPL characterization. Using HSE as a benchmark, the mean relative error (MRE) was calculated and plotted in Figure~\ref{fig:4}f. The SCAN functional performs best with an MRE below $10\%$, followed by r$^2$SCAN at less than $15\%$, while PBE struggles with MRE exceeding $20\%$\cite{51}. On-axis configurations show more systematic behavior, whereas off-axis configurations exhibit greater error variation across defect types. ZPL convergence with respect to k-mesh and supercell size shows consistent trends for all three semi-local functionals (Figure S2), validating the use of $\Gamma$-point calculations. Overall, the SCAN and r$^2$SCAN provide a systematic improvement of ZPL values closer to HSE than PBE, while maintaining computational efficiency comparable to PBE.

\begin{table}
\centering
\caption{Radiative lifetimes of excited-state transitions for NV$^{-1}$-like defects in 4H-SiC (triplet charge states) calculated using SCAN, r$^2$SCAN, and PBE, with HSE results included from references.}
\begin{minipage}{0.8\linewidth}
\begin{ruledtabular}
\begin{tabular}{>{\hspace{0pt}}m{0.052\linewidth}>{\hspace{0pt}}m{0.12\linewidth}>{\hspace{0pt}}m{0.167\linewidth}>{\hspace{0pt}}m{0.11\linewidth}>{\hspace{0pt}}m{0.079\linewidth}} 
\multicolumn{5}{c}{\textbf{Lifetime (ns)}}\\
\textbf{Ion} & \textbf{Conf.}                                                                   & \textbf{SCAN(HSE)}                       & \textbf{r\textsuperscript{2}SCAN} & \textbf{PBE}\cite{14}  \\ 
\hline
\multirow{4}{*}{N}            & \textit{kk}                                                                & 22.99                                    & 23.6                              & 32.2          \\
                                                        & \textit{hh}                                                                & 23.4                                     & 23.4                              & 36.5          \\
                                                        & \textit{kh}                                                                & 18.38                                    & 17.5                              & 32.3          \\
                                                        & \textit{hk}                                                                & 17.82                                    & 18.8                              & 29.4          \\
\hline
\multirow{4}{0.052\linewidth}{O}            & \textit{kk}                                                                & 14.62~(12.5\footnote{\cite{8}})          & 15.2                              & 20.7          \\
                                                        & \textit{hh}                                                                & 13.89                                    & 13.9                              & 20.3          \\
                                                        & \textit{kh}                                                                & 13.28                                    & 12.3                              & 18.1          \\
                                                        & \textit{hk}                                                                & 11.4                                     & 11.6                              & 16.6          \\
\hline
\multirow{4}{0.052\linewidth}{F}            & \textit{kk}                                                                & 13.6                                     & 13.9                              & 69.1          \\
                                                        & \textit{hh}                                                                & 12.8                                     & 13                                & 25.4           \\
                                                        &\textit{kh}                                                                & 12                                       & 11.7                              & 18.7          \\
                                                        & \textit{hk}                                                                & 10.99                                    & 13.4                              & 15.5          \\
\hline
\multirow{4}{0.052\linewidth}{S}            & \textit{kk}                                                                & 136.09                                   & 131.8                             & 182           \\
                                                        & \textit{hh}                                                                & 122.63                                   & 112.3                             & 158           \\
                                                        & \textit{kh}                                                                & 138.07                                   & 114.5                             & 211           \\
                                                        & \textit{hk}                                                                & 105.3                                    & 88.3                              & 162           \\
\hline
\multirow{4}{0.052\linewidth}{Cl}           & \textit{kk}                                                                & 42.91 (30.9\footnote{\label{note:b}\cite{12}}) & 40.1                              & 56.2          \\
                                                        & \textit{hh}                                                                & 38.18 (28.4\text{\textsuperscript{\ref{note:b}}}) & 38.2                              & 49.3          \\
                                                        & \textit{kh}                                                                & 34.96 (28.1\text{\textsuperscript{\ref{note:b}}}) & 36.7                              & 50.0            \\
                                                        & \textit{hk}                                                                & 36.06 (24.3\text{\textsuperscript{\ref{note:b}}}) & 31.0                                & 42.27                                                                                                                                                                         
\end{tabular}
\end{ruledtabular}
\end{minipage}
 \label{tab:2}
\end{table}

The ADAQ workflow calculates the low-temperature optical lifetime $\tau$\cite{44} as:
\begin{equation}\label{eq:lifetime}
    \tau_{i,j} = \frac{3\varepsilon_0 \hbar c^3}{16n\pi^3\nu^3|\vec{\mu}_{i,j}|^2}\text{,}
\end{equation}
where $n$ is the refractive index of the host (2.6473), $\nu$ is the ZPL frequency, and $\mu_{i,j}$ is the transition dipole moment\cite{45}, defined as:
\begin{equation}
    \vec{\mu}_{i,f} = \bra{\phi_f}e\vec{r}\ket{\phi_i}\text{,}
\end{equation}
with $\ket{\phi_i}$ and $\ket{\phi_f}$ as the ground and excited state orbitals. Table~\ref{tab:2} compares radiative lifetimes calculated in the ADAQ workflow for transitions involving the occupied a$_1$(a$^\prime$) state and the lowest unoccupied e(a$^\prime$) level. SCAN and r$^2$SCAN functionals yield radiative lifetimes closest to HSE estimates\cite{8,12}, with shorter lifetimes indicating bright transitions, comparable to the 6.5 ns predicted for the diamond NV center (HSE). Among the defects, SV$^0$ shows consistently higher lifetime estimates across all semi-local functionals. Agreement with HSE improves in the order: PBE $<$ r$^2$SCAN $<$ SCAN.

In summary, we used ADAQ to evaluate the performance of SCAN and r$^2$SCAN meta-GGA functionals for five well-characterized NV-like heteroatom double defect clusters—NV$^{-1}$, OV$^{0}$, SV$^{0}$, FV$^{+1}$, and ClV$^{+1}$—in 4H-SiC. Both Meta-GGAs demonstrate improved accuracy over the semi-local PBE functional in predicting defect properties relevant to quantum technology applications. For neutral defect $E^{q,S}_\text{form}$, the SCAN family bridges the gap between the lower PBE and higher HSE hybrid functional results, with larger deviations observed for later $p$-block elements. Notably, SCAN yields charge transition levels with a MAE of 0.19 eV compared to 0.28 eV for PBE. In optical excitation energy predictions via the $\Delta$SCF method, SCAN achieves ZPL energies closely aligned with HSE, showing MRE below $10$~\%, compared to $15$~\% for r$^2$SCAN and over $20$~\% for PBE. SCAN-based radiative lifetime estimates are similarly closer to HSE results. Overall, the SCAN and r$^2$SCAN functionals are computationally efficient alternatives to HSE, making them suitable for high-throughput screening of new color centers in solids.

See the supplementary material details on elastic properties, electrostatic potentials, ZPL convergence, and the defect hull.

We acknowledge support from the Knut and Alice Wallenberg Foundation (Grant No. 2018.0071). Stretegic Research Area in Material Science on Fucntional Materials at Linköping University, SFO-Mat-LiU No. 2009 00971. J.D. and R.A. acknowledge support from the Swedish Research Council (VR) Grant No. 2022-00276 and No. 2020-05402. I. A. A. is I.A.A Wallenberg Scholar (Grant No. KAW2018.0194). The computations were enabled by resources provided by the National Academic Infrastructure for Supercomputing in Sweden (NAISS) and the Swedish National Infrastructure for Computing (SNIC) at NSC partially funded by the Swedish Research Council through Grant Agreements No. 2022-06725 and No. 2018-05973. We acknowledge PRACE for awarding us access to MareNostrum at Barcelona Supercomputing Center (BSC), Spain. We acknowledge the EuroHPC Joint Undertaking for awarding project access to the EuroHPC supercomputer LUMI, hosted by CSC (Finland) and the LUMI consortium through a EuroHPC Regular Access call.

\noindent
\textbf{AUTHOR DECLARATIONS}

\textbf{Conflict of Interest}

The authors have no conflicts to disclose.

\bibliography{references}

\end{document}


\title{Theoretical characterization of NV-like defects in 4H-SiC using ADAQ with the SCAN and r\textsuperscript{2}SCAN meta-GGA functionals}

\author{Ghulam Abbas}
 \email{ghulam.abbas@liu.se}
\affiliation{Department of Physics, Chemistry and Biology, Link\"oping
  University, SE-581 83 Link\"oping, Sweden}

\author{Oscar Bulancea-Lindvall} 
\affiliation{Department of Physics, Chemistry and Biology, Link\"oping
  University, SE-581 83 Link\"oping, Sweden}

\author{Joel Davidsson} 
\affiliation{Department of Physics, Chemistry and Biology, Link\"oping
  University, SE-581 83 Link\"oping, Sweden}

\author{Rickard Armiento} 
\affiliation{Department of Physics, Chemistry and Biology, Link\"oping
  University, SE-581 83 Link\"oping, Sweden}

\author{Igor A. Abrikosov} 
\affiliation{Department of Physics, Chemistry and Biology, Link\"oping
  University, SE-581 83 Link\"oping, Sweden}
\date{\today}



\date{\today}


\maketitle



%



\section{Supporting Information}
1. Elastic Properties.

2. Average Electrostatic potential. 

3. Zero phonon line convergence tests.

4. Formation energy defect hull calculated with SCAN functional.

5. Formation energy defect hull calculated with r${^2}$SCAN functional.
\newpage

\begin{table}
\centering
\setcounter{table}{0}
    \renewcommand{\tablename}{TABLE.}
    \renewcommand{\thetable}{S\arabic{table}}
\caption{The comparison of Independent elastic constants ($C_{ij}$), Young’s moduli ($E$), bulk moduli ($B$), shear moduli ($G$) calculated with PBE, SCAN, r\textsuperscript{2}SCAN, and HSE exchnage correlation functionals.}
\begin{ruledtabular}
\begin{tabular}{>{\hspace{0pt}}m{0.112\linewidth}>{\hspace{0pt}}m{0.112\linewidth}>{\hspace{0pt}}m{0.112\linewidth}>{\hspace{0pt}}m{0.112\linewidth}>{\hspace{0pt}}m{0.112\linewidth}}
      & \multicolumn{1}{l}{PBE} & \multicolumn{1}{l}{SCAN} & \multicolumn{1}{l}{r2SCAN} & \multicolumn{1}{l}{HSE06}  \\
\hline
$C_{11}$ & 500.3897                & 539.3895                 & 537.3121                   & 542.5782                   \\
$C_{22}$ & 500.4263                & 539.3158                 & 537.6169                   & 542.6106                   \\
$C_{33}$ & 546.2871                & 589.1411                 & 590.2098                   & 593.4174                   \\
$C_{44}$ & 161.6261                & 176.3221                 & 176.3209                   & 175.2578                   \\
$C_{55}$ & 161.4997                & 176.3414                 & 176.0442                   & 175.2837                   \\
$C_{66}$ & 195.1639                & 212.6065                 & 211.9039                   & 211.3141                   \\
$C_{12}$ & 108.7489                & 112.9862                 & 112.7632                   & 119.9963                   \\
$C_{13}$ & 54.6546                 & 53.3901                  & 53.4681                    & 60.4004                    \\
$E$     & 405.198                 & 438.93                   & 439.653                    & 439.774                    \\
$B$    & 216.944                 & 230.004                  & 231.341                    & 237.419                    \\
$G$    & 161.493                 & 176.043                  & 176.302                    & 175.248                   
\end{tabular}
\end{ruledtabular}
 \label{tab:S1}
\end{table}
 
\begin{figure}
    \setcounter{figure}{0}
    \renewcommand{\figurename}{Fig.}
    \renewcommand{\thefigure}{S\arabic{figure}}
     \centering
     \includegraphics[width=1\linewidth]{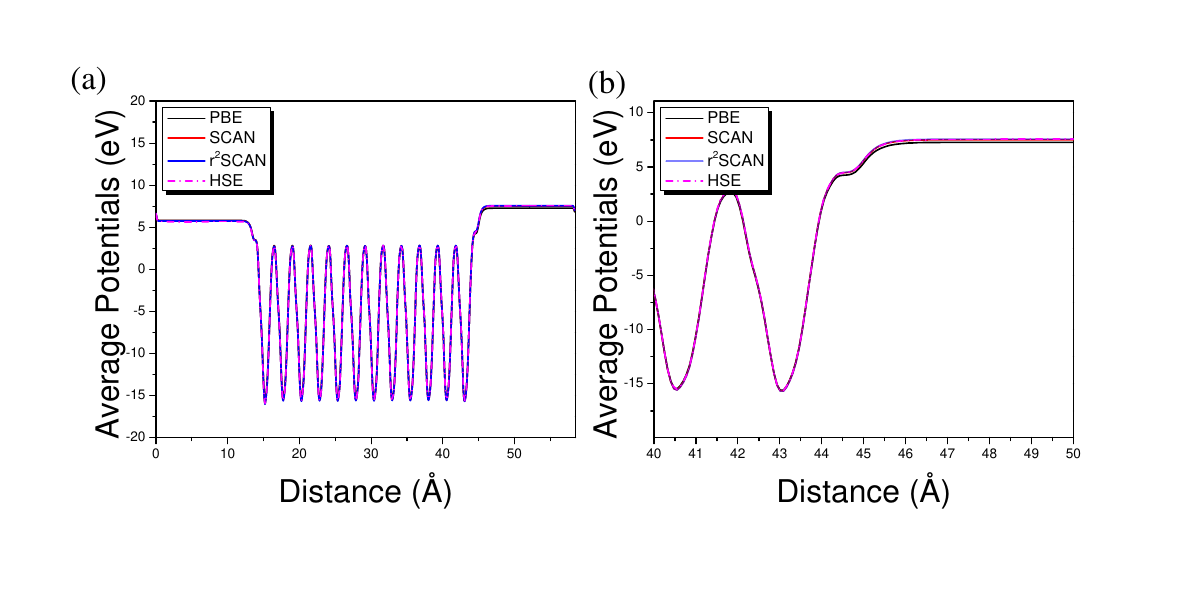}
     \caption{Comparison of average electrostatic potentials (eV) of 4H-SiC slab (0001) model of 12 layers with hydrogen termination as a function of distance (Å) for different exchange-correlation functionals: PBE (black), SCAN (red), r\textsuperscript{2}SCAN (blue), and HSE (magenta). (a) Displays the potential profile over a wider distance range, showcasing the oscillatory behavior within the material region. (b) Zooms in on a smaller region, highlighting the consistency and subtle differences among the functionals near the material interface.}
     \label{fig:S1}
 \end{figure}
 
\begin{figure}
    \setcounter{figure}{0}
    \renewcommand{\figurename}{Fig.}
    \renewcommand{\thefigure}{S2}
     \centering
     \includegraphics[width=0.666\linewidth]{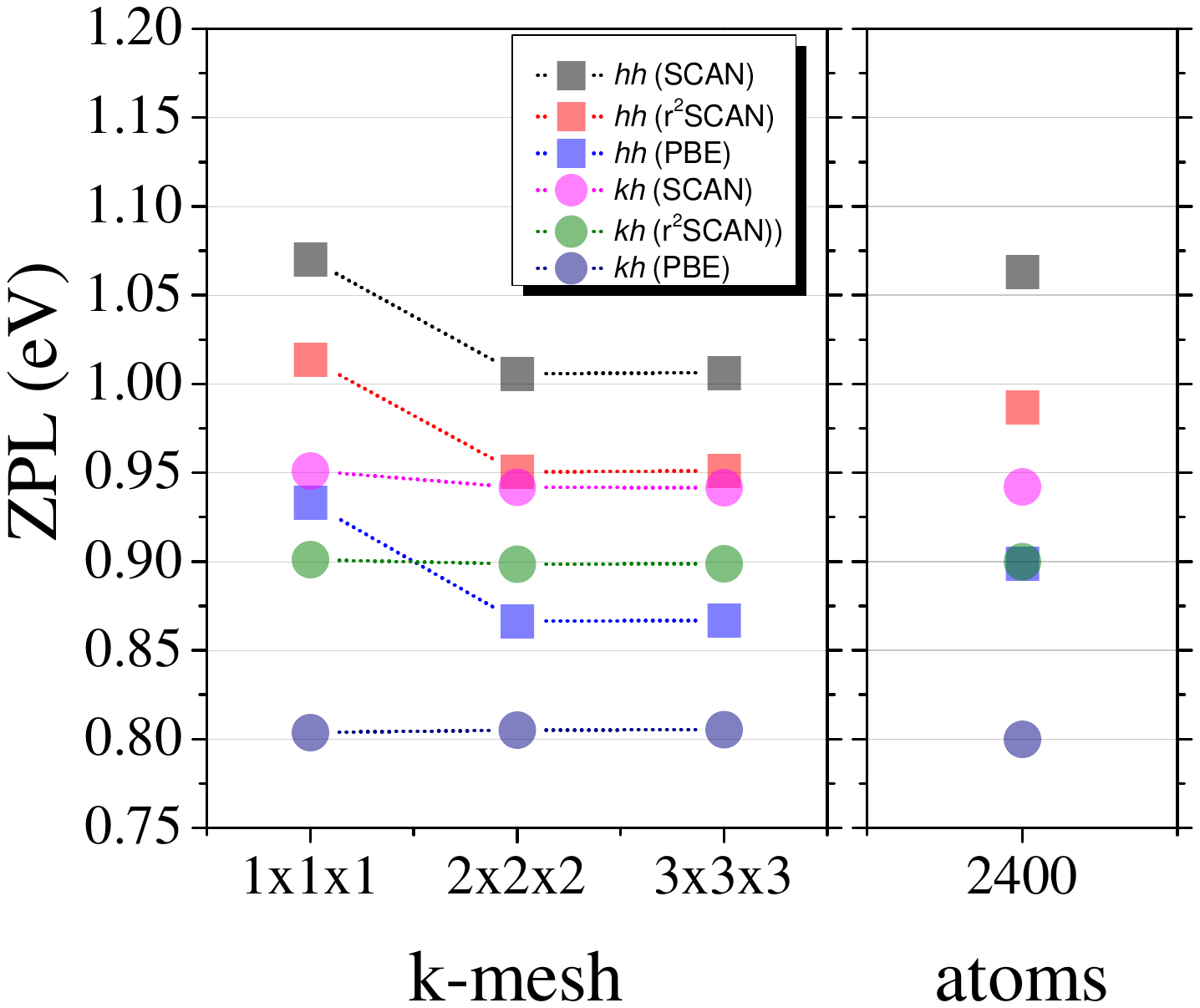}
     \caption{The Zero-phonon line (ZPL) energy convergence of the NV$^{-1}$ defect in 4H-SiC for both hh and kh configurations is shown with respect to the k-point mesh for a 576-atom supercell (left panel) and a 2400-atom supercell at the Gamma point (right panel). The ZPL is computed using three exchange-correlation functionals: SCAN, r\textsuperscript{2}SCAN, and PBE. Overall, negligible k-mesh sensitivity is observed between the different functionals, indicating that Gamma point-based calculations are sufficient for theoretical screening comparisons of the 576-atom supercell imbedded extrinsic defects complex, with minimal variation observed win the 2400-atom system.}
     \label{fig:S2}
\end{figure}

\begin{figure}
    \setcounter{figure}{0}
    \renewcommand{\figurename}{Fig.}
    \renewcommand{\thefigure}{S3}
    \centering
     \includegraphics[width=0.95\linewidth]{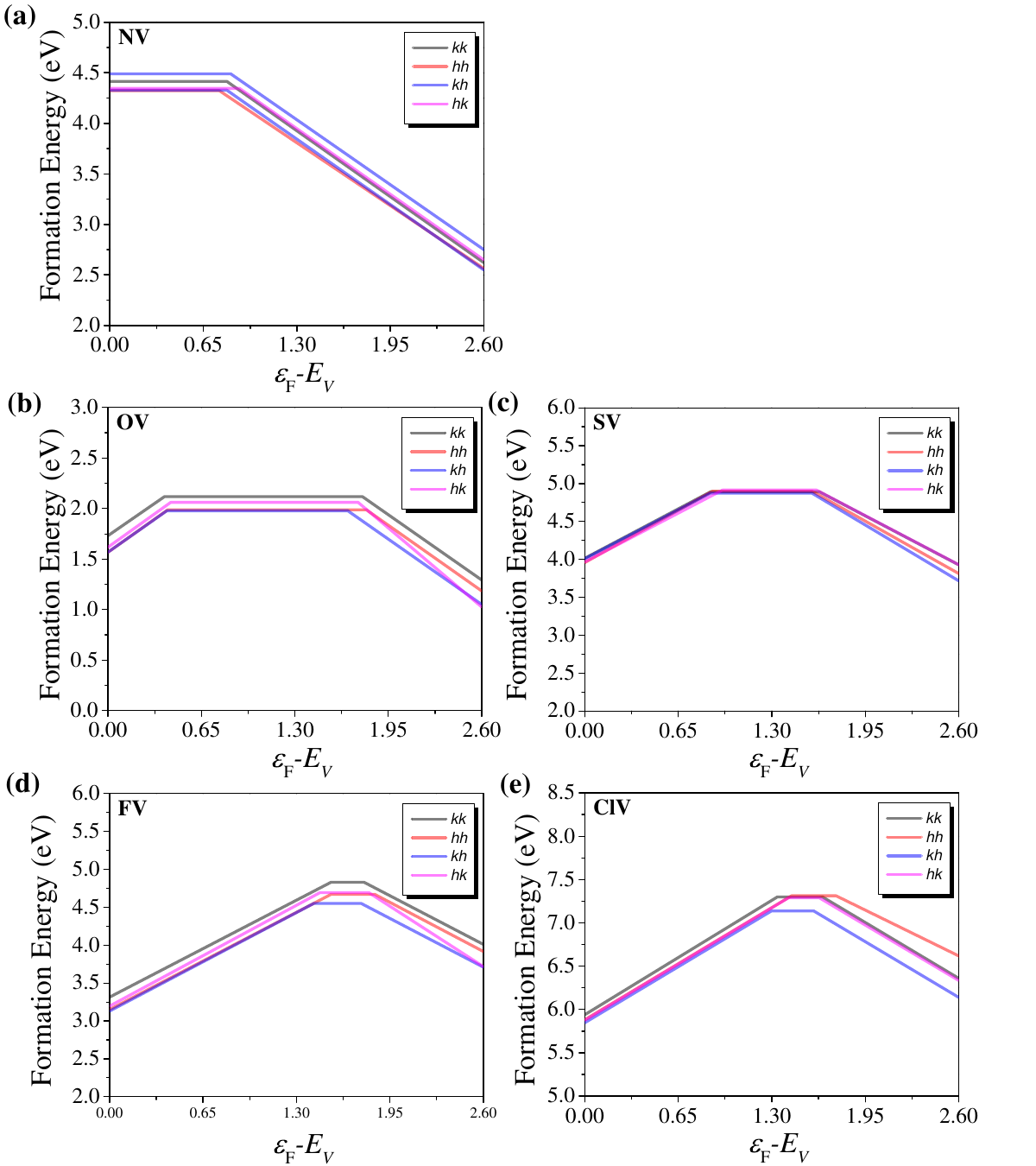}
     \caption{Formation energy (eV) hull as a function of Fermi level relative to the valence band maximum calculated with SCAN functional for (a) NV, (b) OV, (c) SV, (d) FV, and (e) CIV. The configurations are represented by different color codes: \textit{kk} (black), \textit{hh} (red), \textit{kh} (blue), and \textit{hk} (magenta), highlighting the impact on the calculated defect formation energies.}
     \label{fig:S3}
\end{figure}

\begin{figure}
       \setcounter{figure}{0}
    \renewcommand{\figurename}{Fig.}
    \renewcommand{\thefigure}{S4}
     \centering
     \includegraphics[width=0.95\linewidth]{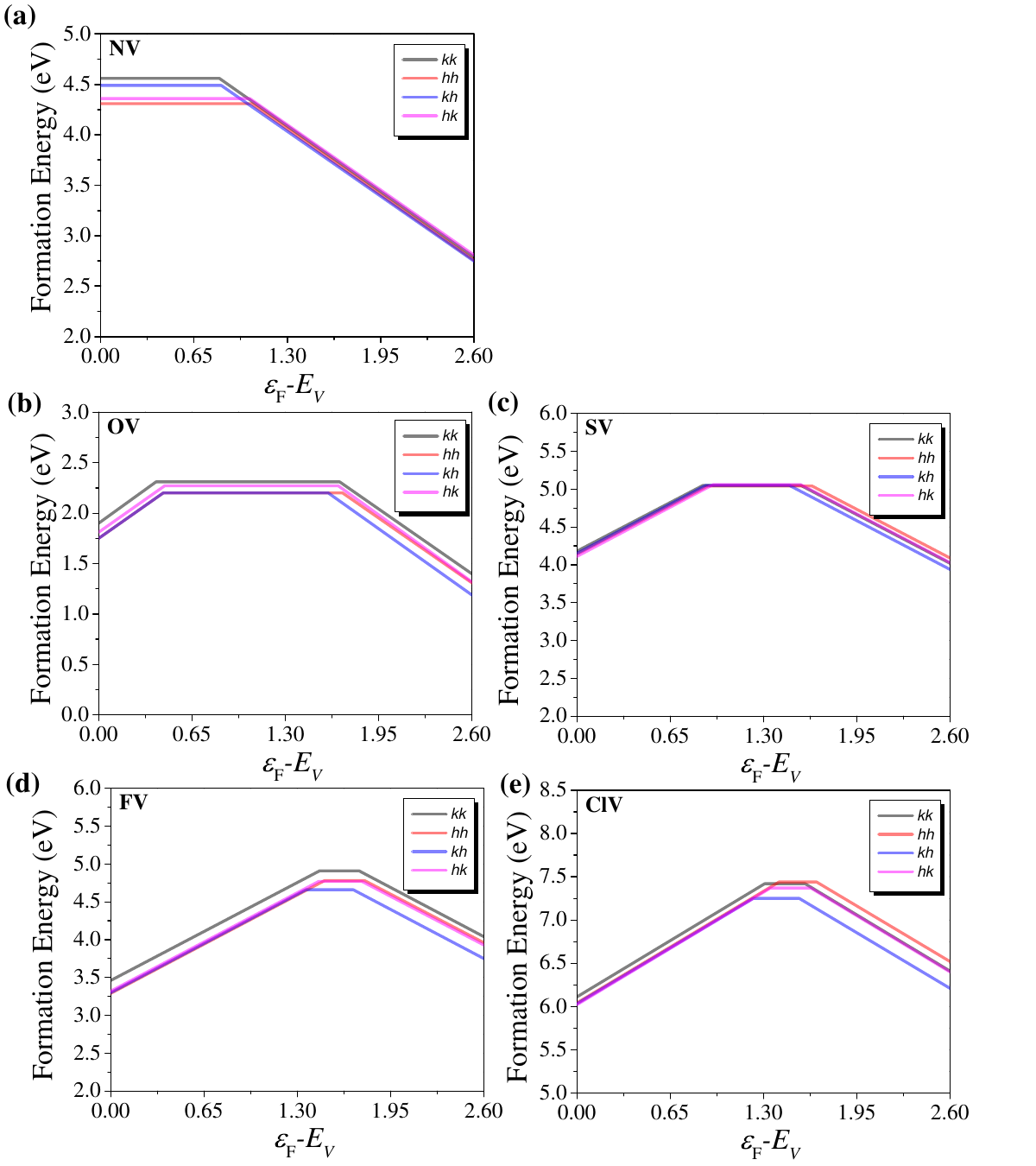}
     \caption{Formation energy (eV) hull as a function of Fermi level relative to the valence band maximum calculated with r$^2$SCAN functional for (a) NV, (b) OV, (c) SV, (d) FV, and (e) CIV. The configurations are represented by different color codes: \textit{kk} (black), \textit{hh} (red), \textit{kh} (blue), and \textit{hk} (magenta), highlighting the impact on the calculated defect formation energies.}
     \label{fig:S4}
\end{figure}


%


